\documentstyle[aps]{revtex}
\input{psfig}
\title{Baryons As Hyperspherical O(4) Partial Waves --\\
Is This The Message From The Spectra?}
\author{M. Kirchbach 
\footnote{ and {\it Institute for Nuclear Physics, University Mainz,
D-55099 Mainz, Germany } }\\
{\it Escuela de Fisica, Univ.\  Aut.\ de Zacatecas,}\\
{\it Apartado  Postal  C-580, Zacatecas, ZAC 98068 Mexico} }

\begin{document}
\maketitle
\begin{abstract}
It is argued that the baryon excitations group to four-dimensional partial 
waves described by means of the three Rarita--Schwinger (RS) fields 
$\lbrace {{\sigma -1}\over 2},{{\sigma -1}\over 2}\rbrace \otimes \lbrack 
\lbrace {1\over 2}, 0\rbrace \oplus \lbrace 0,{1\over 2}\rbrace \rbrack $
with  $\sigma =2,4$ and $6$, where all components happen to be occupied. In 
the O(4) decomposition of the $\pi N$ scattering amplitudes, the RS spin- 
and parity clusters appear as poles on the complex energy plane, socalled
H\"ohler poles. This phenomenon indicates that the symmetry of the $\pi N$ 
scattering amplitude is O(4) and thereby the  space--time version of chiral 
symmetry, rather than O(3). Accordingly, the baryon spectrum generating 
algebra is su(2)$_I\otimes $su(3)$_c\otimes$o(1,3)$_{ls}$ rather than 
su(6)$_{sf}\otimes $su(3)$_c\otimes $o(3)$_l$. The nucleon and $\Delta $
spectra below $\sim $2500 MeV are complete up to only 5 `missing' 
resonances. The three O(4) poles are distributed over two distinct Fock 
spaces of opposite vacuum parities thus defining the energy scale of the 
chiral phase transition for baryons. Within this new symmetry scenario, the 
covariant description of the RS baryon clusters is straightforward and their 
averaged masses are fitted by a Balmer-series like formula emerging from a 
simple quark-diquark model in the O(4) basis with Coulomb potential and a 
four-dimensional rigid rotator.

\end{abstract}

PACS:\qquad 14.20.Gk, 11.30.Ly, 03.65.Pm, 11.30.Rd 

Key words: Higher--spin baryons, Rarita-Schwinger fields, quark-diquark model,
           chiral phase transition

\vspace*{0.25truein}

\noindent
One of the most important concepts of baryon spectroscopy is the 
classification of the resonances according to the representations of 
the non--relativistic group SU(6)$_{sf}\otimes $O(3)$_l$.
Through this group the trivial spin-flavor $(sf)$ correlation
between three quarks in the 1s-shell has been assumed to apply
to arbitrary internal orbital angular momenta ($l$).
In doing so, states like, say, the positive parity resonances
P$_{13}$(1720), F$_{15}$(1680), F$_{35}$(1905), and F$_{37}$(1950),
are viewed to belong to a 56(2$^+$)--plet, the 
P$_{11}$(1710) excitation is treated as a member 
of a 70(0$^+$)--plet, while the negative parity baryons
S$_{11}$(1535), D$_{13}$(1520), S$_{11}$(1650), D$_{13}$(1700), and
D$_{15}$(1675) are assigned to a 70(1$^-$)--plet. The above examples 
clearly illustrate 
how states separated by only few MeV, like the D$_{15}$(1675), 
F$_{15}$(1680), and P$_{11}$(1700) states, are distributed over
three different SU(6)$_{sf}\otimes $O(3)$_l$ representations, whereas,
on the other hand, resonances separated by about 200 MeV like the 
D$_{13}$(1520) and the D$_{13}$(1700) ones, 
are assigned to the same multiplet \cite{Part,Bhaduri}. 
The basic idea of the 
multiplets as well separated families of particles of 
different internal but identical space--time properties, 
appears quite inappropriate here, where 
the spacing between the SU(6)$_{sf}\otimes $O(3)$_l$ multiplets is
much smaller as compared to the maximal mass splitting within them.
In addition, the SU(6)$_{sf}\otimes $O(3)$_l$ baryon
classification scheme predicts a substantial excess of resonances, 
called `missing', which yet have not been observed.
Nonetheless, the SU(6)$_{sf}\otimes $O(3)$_l$ symmetry predictions
on the mass spectrum have been considered as quite
satisfactory so far, with the excuse that the deviations from the observed 
masses of about $\pm $150 MeV are small on the scale of 1500-2500 MeV.

On the other side, speed plot analysis of the pole positions 
on the complex energy plane of various baryon resonances  
($L_{2I, 2J}$) with masses below $\sim $ 2500 MeV performed by 
H\"ohler and Sabba-Stefanescu~\cite{Hoehler} have revealed a 
well-pronounced partial-wave clustering in baryon spectra, socalled 
H\"ohler poles. As a representative example, the grouping of the 
S$_{11}$, P$_{11}$, P$_{13}$, D$_{13}$, D$_{15}$ and F$_{15}$ states around 
the pole (1665$\pm$ 25) -i (55$\pm $15) may be mentioned. This is quite a 
surprising result as it was not anticipated by any hadron model. In view of 
the H\"ohler clustering, it appears timely to question the 
SU(6)$_{sf}\otimes $ O(3)$_l$ classification and search for a new 
scheme for baryons which matches better with the observed spin-and parity 
grouping of the excited states and contains a much smaller
number of unobserved (`missing') resonances. 

To the best of our knowledge, the problem of H\"ohler's poles
was challenged only recently in a series of papers \cite{Ki97-98a} where
it was shown that they can be identified in a natural way
with four-dimensional hyperspherical O(4) partial waves, here denoted by 
$\sigma _{2I , \eta }$, with $ \sigma =2,4$, and $6$, and $\eta =\pm 1$ 
(see Fig. 1). This means that nature seems to favor the O(4) partial wave 
decomposition of the $\pi N$ scattering amplitude over the O(3) one.
The O(4) partial waves from above are well known from the Coulomb problem, 
where they correspond to the (even) principal quantum numbers $n =\sigma $. 
They join (approximately) mass degenerate O(3) states of integer internal 
angular momenta, $l$, with $l=0,..., \sigma -1 $. All O(3) partial waves,
$\sigma_{2I,\eta ;lm}$, contributing to a given O(4) pole, 
have either natural ($\eta =+1$), or unnatural ($\eta = -1$) parities. 
In other words, they transform with respect to the space inversion
 operation ${\cal P}$ as
\begin{eqnarray}
{\cal P} \sigma _{2I, \eta ;l m } &=& \eta e^{i\pi \, l}\, 
\sigma _{2I, \eta ;l -m } \, \nonumber\\
l=0^\eta ,1^{-\eta } ,..., (\sigma -1) ^{-\eta }  \, , 
&\qquad & m=-l,..., l\, .
\label{party}
\end{eqnarray}
In coupling a Dirac spinor to the O(4) multiplets from above, the
spin ($J$) and parity $(\pi $) quantum numbers of the baryon resonances are
created as 
\begin{eqnarray}
J^\pi &=& {1\over 2}^\eta , {1\over 2}^{-\eta }, {3\over 2}^{-\eta }, ...,
(\sigma -{1\over 2}) ^{-\eta } \, .
\label{coupl_scheme}
\end{eqnarray}
The first four-dimensional hyperspherical partial wave is always 
$ 2_{2I, +}$. From Eqs.~(\ref{party}) and (\ref{coupl_scheme}) 
follows that it unites the first spin- ${1\over 2}^+$, ${1\over 2}^-$, and 
${3\over 2}^-$ resonances. Indeed, the relative $\pi N$ momentum $L$
takes for $l=0^+$ the value $L=1^+$ and corresponds to the P$_{2I, 1}$ state,
while for $l=1^-$ it takes the two values $L=0^-$, and $L=2^-$
describing in turn the S$_{2I, 1}$ and D$_{2I, 3}$ resonances. 
The isospin quantum number $I$ takes the three different values 
$I=1/2, 3/2$, and $0$. The natural parity of the first 
O(4) partial waves reflects the arbitrary selection of a scalar vacuum through 
the spontaneous breaking of chiral symmetry. Therefore, up to the three lowest 
$N, \Delta$, and $\Lambda $ excitations,  chiral symmetry is still in the 
Nambu--Goldstone mode. All the remaining non--strange baryon resonances have 
been shown in \cite{Ki97-98a} to belong to either $4_{2I, -}$, or $6_{2I, -}$.
{}For example, one finds all the seven $\Delta $--baryon resonances 
$S_{31}, P_{31},P_{33}, D_{33},D_{35}, F_{35}$ and $F_{37}$
from the $4_{3, -}$ partial wave to be squeezed within the 
narrow mass region from 1900 MeV to 1950 MeV, while the I=1/2 resonances 
paralleling them, of which only the $F_{17}$ state is still 'missing' from 
the data, are located around 1700$^{+20}_{-50}$ MeV. Therefore, the F$_{17}$ 
resonance is the only non--strange state with a mass below 2000 MeV which is 
'missing' in the new scheme. 
This one `missing' resonance has to be compared to at least 10 resonances 
considered as 'missing' within the traditional SU(6)$_{sf}\otimes $O(3)$_l$ 
schemes (see \cite{Part,Bhaduri} for reviews).
In continuing by paralleling baryons from the third nucleon and 
$\Delta $ clusters with $\sigma $=6, one finds in addition the four
states H$_{1, 11}$, P$_{31}$, P$_{33}$, and D$_{33}$ with masses above
2000 MeV to be `missing' from the new scheme.
The H$_{1, 11}$ state is needed to parallel the well established
H$_{3, 11}$ baryon, while the $\Delta $-states P$_{31}$, P$_{33}$, and
D$_{33}$ are required as partners to the (less established) P$_{11}$(2100),
P$_{13}$(1900), and D$_{13}$(2080) nucleon resonances.
The second and third non--strange baryon clusters have been shown
in \cite{Ki97-98a} to be built upon a pseudoscalar vacuum and are of 
unnatural parities. The latter circumstance signals chiral symmetry 
restoration here and fixes the mass scale of the chiral phase transitions 
for baryons. It is remarkable, that the  approximately equidistant 
cluster spacing of about 200 MeV to 300 MeV
between the mass centers of the O(4) multiplets appearing now,  
is by a factor 3 to 6 larger as compared, for example,
to the maximal mass splitting of 50-70 MeV within
the $2_{1,+}$, $2_{3,+}$, $4_{1,-}$, and $4_{3,-}$ partial waves
(see Table 1). 

The degeneracy of the states $({1\over 2}^-$- ${3\over 2}^-)$, and
$({3\over 2}^+$-${5\over 2}^+)$ from the region around 1700 MeV was noticed in
\cite{Hosaka} where it was explained by means of a quark model with a deformed
harmonic oscillator, denoted by DOQ. However, several states from that region 
like the ${1\over 2}^+$ and ${5\over 2 }^-$ ones, don't fit into
DOQ scheme and have been left out of consideration. 
We here stress that the degeneracy of the resonances considered in \cite{Hosaka}
is the direct consequence of their belonging to the O(4) clusters.
The big advantage of our classification scheme 
is that there are no states out of the O(4) partial wave systematics.

To illustrate the substantial reduction of the number of the
'missing' resonances within the O(4) 
classification scheme of the internal orbital angular
momenta in \cite{Ki97-98a}, it is quite instructive
to consider as an example how the quantum numbers of the three
lowest baryon excitation P$_{2I\,  1}$, S$_{2I\,  1}$, and D$_{2I\, 3}$,
can emerge from the minimal O(4) symmetric quark-diquark configuration space 
spanned by the $1s$, $1p$, and $2s$-- single--particle shells 
with an approximate $1p-2s$ degeneracy.
The one-particle--one-hole configurations in this
space give rise to the following orbitally excited (active)
diquarks (in standard shell--model notations) of both
natural and unnatural parities :
\begin{equation}
\lbrack 1s_{{1\over 2}}^{-1}\otimes  2s_{{1\over 2}}^1
\rbrack ^ {l=0^+,1^+}\, ,
\qquad 
\lbrack 1s_{{1\over 2}}^{-1} \otimes 1p_{{1\over 2};{3\over 2}}^1)
\rbrack ^{l=0^-,1^-, 2^-}\, ,
\label{1P_1H}
\end{equation}
Note that we consider the quark-diquark model in the $j$-$j$ rather
that in the $LS$ coupling exploited in the traditional 
SU(6)$\otimes $O(3) quark models.

Now, the quantum numbers of the first P$_{2I\, , 1}$, 
S$_{2I\, , 1}$ and D$_{2I\, , 3}$ excitations 
are determined through the coupling of the spectator
$1s_{{1\over 2}}$ quark to natural parity diquarks,
such like
\begin{eqnarray}
\left( \lbrack 1s^{-1}_{{1\over 2}} \otimes 
2s^1_{{1\over 2}} \rbrack^ {l=0^+} \otimes
1s_{{1\over 2}}^1\right)^{{1\over 2}^+}\, ,
 &\quad & \left(\lbrack 1s^{-1}_{{1\over 2}} \otimes 
1p^1_{{1\over 2}} \rbrack^{ l=1^-}\otimes
1s_{{1\over 2}}^1\right)^{{1\over 2}^-;{3\over 2}^-}\, .
\label{natural_par}
\end{eqnarray}
None of the remaining unnatural parity configurations 
\begin{eqnarray}
\left(\lbrack 1s^{-1}_{{1\over 2}}\otimes 
1p^1_{{1\over 2};{3\over 2}}\rbrack ^{l=0^-, 2^-}\otimes 
1s^1_{{1\over 2}}\right)^{{1\over 2}^-, {3\over 2}^-, {5\over 2}^-}\, ,
 &\quad & \left(\lbrack 1s^{-1}_{{1\over 2}}\otimes 
2s^1_{{1\over 2}}\rbrack ^{l=1^+}\otimes 
1s^1_{{1\over 2}}\right)^{{3\over 2}^+}\, ,
\label{unnat_par}
\end{eqnarray}
has been observed in the mass region around 1500 MeV so far. 
These configurations rather occur as members of the unnatural parity
$4_{2I,-}$ and $6_{2I, -}$ clusters.    
Such a radical truncation of the 3q--configuration space
is understandable provided, to some approximation, the diquarks 

\, i) behave as pointlike bosonic subbaryon degrees of freedom,

ii) their parities are selected to be either all natural, or all unnatural.

To satisfy the parity selection rule one has to assume that the diquarks
are created as one--particle states within a given Fock space 
(to be denoted by ${\cal F}$) built upon a vacuum
(denoted by $|0^\eta \rangle $) of either positive ($\eta =+$), 
or negative $(\eta = -$) parity. For example, the particle-hole creation 
operators $A_{\sigma \eta ;lm}^\dagger $, with $l=0^+, 1^-$,
which describe in turn the scalar and vector diquarks from the O(4) partial 
wave $ 2 _{1, + } $, 
\begin{eqnarray}
A^\dagger _{2_{1,+}  ;lm} 
& =& 
\sum_{m_1m_2} 
(-1)^{{1\over 2}-m_2}({1\over 2} m_1 {1\over 2} \, -m_2|lm)
b^\dagger _{2_{1,+}; 2s_{{1\over 2}} m_1 }
b_{2_{1,+}; 1s_{{1\over 2}} m_2}\, ,
\label{orb_excit}
\end{eqnarray}
have to act onto the baryon ground state, $|(1s_{1\over 2})^3\rangle $,
in the same way as a fundamental bosonic operator, here denoted by 
$a^\dagger_{2_{1,+};lm}$, acts onto its Fock vacuum, so that the following 
mappings hold:
\begin{eqnarray}
A^\dagger _{2_{1,+}; lm}\,
|(1s)^3\rangle \simeq  a^\dagger_{2_{1,+} ;lm}\, |0^+\rangle \, ,
&\quad &{\cal P} a^\dagger _{2_{1,+} ;lm} |0^+\rangle  = 
 e^{i\pi l}a^\dagger _{2_{1,+} ; l,\,  -m}\,|0^+\rangle \,  .
\label{p-h_oper}
\end{eqnarray}
The mapping in Eq.~(\ref{p-h_oper}) is nothing but the expression
for the {\bf superselection\/} rule from above allowing only  
{\bf diquarks having all either natural or unnatural parity and grouped to
O(4) families \/}, called hyperquarks (HQ) in the following,
as relevant degrees of freedom for the structure of the excited baryons. 
From this stage on we will ignore the constituent character
of the O(4) boson operators and consider in the following the hyperquarks
as  {\it fundamental \/} degrees of freedom. 
The idea of the pointlike character of the diquarks 
has been exploited in the literature to reduce
the three-quark  Faddeev equations to a two-body quark-diquark
Bethe-Salpeter equation (see, for example \cite{Hel97,Kus}).
The essential difference between the present quark-hyperquark model (QHM)
and the customary quark-diquark models (QDM) (see \cite{Anselmino}
for a digest) is the assumed O(4) clustering of the diquarks and the
parity selection rule leading to the observed clustering of baryons.
To be specific, the operator D$^\dagger_{2_{1,+} } $ which creates
the lightest hyperquark is defined as
the following linear combination of fundamental one-boson states
\begin{eqnarray}
 D^\dagger_{2_{1,+} }|0^+ \rangle  =
\sum_{lm}\, c_{lm}  a^\dagger _{2_{1,+} ;lm }\,|0^+\rangle \,  ,
&\quad & \sum_{lm}|c_{lm}|^2 = 1\, ,
\nonumber\\
\langle 0^+|  D^\dagger_{2_{1,+} }|0^+\rangle  
 &=& \sum_{lm}\, c_{lm}\,  {\cal R}_{2 l} (r) 
Y_{2 lm}(\alpha ,\theta ,\phi) \, .
\label{hyperquark}
\end{eqnarray}
In Eq.~(\ref{hyperquark}) the radial part of the hyperquark 
wave function has been denoted by  ${\cal R}_{\sigma l}(r)$,
while its angular part has been determined by
the four-dimensional hyperspherical harmonics
$Y_{\sigma lm}$ defined in the standard \cite{Shibuya} as
\begin{eqnarray}
Y_{\sigma lm}(\alpha ,\theta ,\phi)\, 
 &=&
 i^{\sigma-1-l} 2^{l+1}l!\, 
 {{\sigma (\sigma -l-1)}\over {2\pi (\sigma +1)}} \, 
\sin^\sigma \alpha \, {\cal C}^{l+1}_{\sigma -l-1} (\cos \alpha )
\, Y^l_m (\theta, \phi )\, .
\label{Gegenb}
\end{eqnarray}
Here, ${\cal C}^{l+1}_{\sigma -l-1} (\cos \alpha )$ denote the 
Gegenbauer polynomials, while Y$^l_m(\theta ,\phi )$ are the 
standard three-dimensional spherical harmonics.

In general, a Lorentz covariant spin- and parity 
cluster $ \sigma_{1,\eta} $ is now described as
\begin{eqnarray}
\sigma_{1, \eta}  &=& 
D^\dagger_{\sigma _{1,\eta} }\, b^\dagger_{1s_{ {1\over 2} } }\,
|0^\eta \rangle \,  \lbrack T^1\otimes \chi^{{1\over 2}}\rbrack^{{1\over 2}}  \, , 
\label{island}
\end{eqnarray}
where $T^1$ stands for the flavor part of the wave function of the (nonstrange) 
hyperquark as a {\bf symmetric\/} isovector state, while 
$\chi^{ {1\over 2}} $ is the ordinary isospinor of the third quark. 
It will become clear in due course that the restriction to 
isovector hyperquarks in Eq.~(\ref{hyperquark}) ensures that the 
complete space-time-flavor-color wave function of the $I={1\over 2}$ 
clusters is totally antisymmetric, as it should be in order to respect
the Fermi-Dirac statistics for quarks.

The O(4) symmetry ansatz for the quark-hyperquark model
assumed  in the present work is independently supported by the observed 
rapid convergency of the covariant diquark models in the basis of
the Gegenbauer polynomials considered among others in \cite{Hel97,Kus}. It is 
worthy of being pursued especially because of the quite uncertain experimental 
status of the 'missing' resonances. The relevant spectrum generating algebra 
deduced in \cite{Ki97-98a} is 
\begin{equation}
su(2)_I\otimes su(3)_c\otimes o(1,3)_{ls}\, ,
\label{Lor_sp}
\end{equation}
and the baryon structure acquires features similar to those
of the hydrogen atom.
 
The apparent analogy between the spectrum of the hydrogen atom and 
the baryon spectra raises the question whether the positions of the Lorentz 
covariant spin--clusters is determined by the inverse squared $1/\sigma ^2$,
the analogue of the inverse squared of the principle quantum number of the 
Coulomb problem, and follow a type of Balmer-series like pattern. The answer 
to this question is positive. Below we give a simple empirical recursive 
relation which describes with quite an amazing accuracy the reported mass 
averages of the resonances from the Lorentz multiplets  (see Table 1) with 
$\sigma =2, 4$, and $6$ only in terms of the O(4) quantum number $\sigma $, on 
the one side, and the two mass parameters $m_1$=600 MeV, and $m_2$=70 MeV, on 
the other side,
\begin{eqnarray}
M_{\sigma '}  -M_{\sigma } &=&
m_1 \, \left( {1\over {\sigma^2 }}-{1\over {(\sigma \, ')\,\,  ^2}}\right) 
+{1\over 2} m_2\, \left( {{\sigma '^2-1}\over 2}-
{{\sigma^2 -1}\over 2}\right)  \, .  
\label{Balmer_ser}
\end{eqnarray}
The first term on the r.h.s. in Eq.~(\ref{Balmer_ser}) is
the typical difference between the energies of two single particle states
of principal quantum numbers $\sigma $, and $\sigma '$, respectively,
occupied by a diquark with  mass  $m_1=600 $ MeV moving in a Coulomb potential.
To explain the origin of the second term one needs to remember that
the so(4) algebra, in being six dimensional,
is the direct sum of two independent 
three-dimensional right- (R) and left (L) handed su(2) algebras,
\begin{equation}
so(4)=su(2)_L\oplus su(2)_R\, .
\label{o4_env}
\end{equation}
Therefore, the purely space-time chiral group SU(2)$_L\otimes $SU(2)$_R$
acts as the universal covering of SO(4), and the irreducible SO(4) 
representations can be labeled by two SU(2) indices, denoted by
 $j_1$, and $j_2$ here, according to  $\lbrace j_1,j_2\rbrace $.
{}For this reason, the O(4) multiplets $\lbrace j_1,j_2\rbrace $ are 
eigenstates of the sum of the Casimir operators $J_1^2$, and $J_2^2$ 
associated in turn with the two SU(2) groups from above, and one finds
\cite{deAlf}
\begin{equation}
(J_1^2 +J^2_2)\lbrace j_1,j_2\rbrace = 
\left( j_1(j_1+1) +j_2(j_2+1)\right) \lbrace j_1,j_2\rbrace\, . 
\label{so4_casimir}
\end{equation}
{}From this point of view, the second term on 
the r.h.s in Eq.~(\ref{Balmer_ser}) emerges as the eigenvalue of 
the O(4) partial wave considered with respect to the direct sum
of the two three-dimensional rotators from Eq.~(\ref{so4_casimir})
and is determined by 
\begin{eqnarray}
\left({ {1   } \over {2{\cal J}} }J_1^2 +
{{1} \over {2{\cal J}} }J_2^2\right) 
 \lbrace j_1, j_2\rbrace 
&=& { {1}\over {2{\cal J}} }\, \left( j_1(j_1+1) +j_2(j_2 +1)\right)
\lbrace j_1, j_2 \rbrace \, ,\nonumber\\
\mbox{for} \quad j_1=j_2= { {\sigma -1}\over 2 }\, ,
&\quad &  \quad j_1(j_1+1) +j_2(j_2+1) = {{\sigma^2-1} \over 2}\, . 
\label{two_rotators}
\end{eqnarray}
The term  ${{\sigma^2-1}\over 2}$ in Eq.~(\ref{Balmer_ser}) is the 
generalization of the three-dimensional j(j+1) rule to four dimensions.
In Eq.~(\ref{two_rotators})  ${\cal J}$ plays the role of an
effective inertial moment.
Comparison of Eq.~(\ref{two_rotators}) to (\ref{Balmer_ser})
reveals that the parameter 1/$m_2$ =2,82 fm 
corresponds to the inertial moment ${\cal J}=2/5 MR^2$ of some
`effective'  rigid-body resonances with mass $M = 1085$ MeV and a radius 
R=1,13 fm. 
Therefore, the energy spectrum in Eq.~(\ref{Balmer_ser}) can be considered to 
emerge from a quark-hyperquark model with a Coulomb potential 
(${\cal H}_{Coul}$) and a four--dimensional rigid rotator ($T_{rot}^{(4)}$). 
The corresponding Hamiltonian ${\cal H}^{QHM}$ is given by
\begin{eqnarray}
{\cal H}^{QHM}  &=&  {\cal H}_{Coul}+T_{rot}^{(4)}\nonumber\\
&=&{\alpha_C\over r} + {{1 }\over {2{\cal J}}} (J_1^2 +J_2^2)\, .
\label{Hamlit}
\end{eqnarray}
This Hamiltonian is diagonal in the basis of the O(4) partial waves
and the parameter $m_1/\alpha_C$ plays a role similar to that of the 
Rydberg constant. Note that while the splitting between the Coulomb 
states decreases with increasing principal quantum number $\sigma $,
the difference between the energies of the rotational states increases 
linearly with $\sigma $ so that the net effect is an 
approximate equidistancy of  the baryon cluster positions.
In extending the Hamiltonian in Eq.~(\ref{Hamlit}) to include
O(4) violating terms such like $\sim l\cdot s$,  $\sim l^2 $,
or introducing different inertial momenta ${\cal J}_i$ to account
for possible deformation effects, 
the O(3) splitting of the O(4) clusters can be studied along the line
of the collective models of nuclear structure \cite{EisenbGr}.
{}Finally, the purely space-time version of chiral symmetry of our model 
in Eq.~(\ref{Lor_sp}) can also be extended to include the 
combined space-time \& flavor chiral symmetry leading to
a Goldstone-boson-quark interaction in the spirit of Refs.~\cite{DOR}.

The introduction of  O(4) correlations between the diquark O(3) partial 
waves brings numerous advantages over treating them as independent ordinary 
spherical partial waves. In particular, it allows for the relativistic 
description of the O(4) hyperquark  propagators and therefore, for the 
relativistic propagators of the resulting baryon clusters. 
Indeed, baryons grouped to O(4) partial wave--clusters $\sigma_{2I, \eta }$ 
of momentum $p_\mu $, and mass $M$ are nothing but
the reducible Lorentz representations 
$\lbrace {{\sigma -1}\over 2 },{{\sigma -1}\over 2 }\rbrace
\otimes \lbrack \lbrace {1\over 2},0\rbrace \oplus 
\lbrace 0, {1\over 2}\rbrace\rbrack $
known as Rarita--Schwinger (RS) spinors.
They are described by {\bf totally symmetric} traceless rank--$(\sigma -1 )$ 
Lorentz tensors  with Dirac spinor components and  
satisfy both the Dirac and Proca equations:
\begin{eqnarray}
(p\cdot \gamma -M)\Psi_{\mu_1 \mu_2...\mu_{\sigma -1 }} &=& 0\, ,\\
(g^{\nu \mu } -{1\over M^2}p^\nu p^{\mu_1}) 
\Psi_{\mu_1 \mu_2...\mu_{\sigma -1 }} &=& 
\Psi^\nu_{\,\,\, \mu_2...\mu_{\sigma -1}}\, .
\label{Dirac_Proca}
\end{eqnarray}
The RS spinors have been used by Weinberg in his
classical work \cite{J00J} for embedding the higher--spin states 
$J=\sigma - {1\over 2}$. 
The essential difference between Weinberg's scheme and the one presented here 
is that the lower-spin states entering the RS spinors are 
{\bf no longer redundant\/} components that need be eliminated, but 
{\bf physically observable O(3) resonances\/} reflecting the
composite character of baryons. 
The totally antisymmetric character of the $I={1\over 2}$ cluster wave function 
is now ensured by the ansatz in Eq.~(\ref{island}) where the hyperquark was 
constructed as an isovector. As both the space-time and isospin parts of wave 
function of the $\sigma_{1,\eta }$ clusters are now totally symmetric by 
construction, its color part is, as usual, in the totally antisymmetric color 
singlet state. This structure of the cluster wave function is fully consistent 
with the symmetry in Eq.~(\ref{Lor_sp}), where the SU(3)$_c$ transformations
are completely independent from the flavor and space-time ones.
In such a case the diquark correlation in isospin space
does not necessarily imply  a similar correlation in color space. For this reason, 
in color space the three quarks can still be treated as bound to an 
antisymmetric singlet. Moreover, the space-time hyperquark correlation does not 
necessarily require a diquark in  isospin space. This allows one to construct the 
correct wave function for the $I={3\over 2}$ clusters in considering  hyperquarks 
as purely space--time objects, while keeping in both isospin and flavor
spaces the concept of the three independent quarks bound to a symmetric 
isospin- and a totally antisymmetric color states, respectively.

The Lorentz-covariant hyperquark propagator 
$D^{HQ} _{\mu_1 \mu_2...\mu_{\sigma-1} ;\nu_1 \nu_2...\nu_{\sigma-1}} $ 
is easily constructed from Proca's spin-1 projectors, 
on the one side, and the mass-shell condition, on the other side, and 
is given by
\begin{equation}
D^{HQ}_{\mu_1 \mu_2...\mu_{\sigma -1};\nu_1\nu_2...\nu_{\sigma-1} } =
{{ 
\bigotimes_{n=1}^{n=\sigma -1 } 
(g_{\mu_n\nu_n} - {1\over M^2}  p_{\mu_n}p_{\nu_n}) }
\over {p^2-M^2} }\, .
\label{hquark_prop}
\end{equation}
The relativistic propagators 
of the spin-parity baryon clusters are obtained as direct products of
the hyperquark propagator in Eq.~(\ref{hquark_prop})
and the Dirac projector corresponding to the spectator quark
\begin{equation}
S_{\mu_1\mu_2...\mu_{\sigma -1 }\,  ;\, \nu_1\nu_2 ...\nu_{\sigma -1}}=
{ { \gamma\cdot\, p  +M }\over {2M} }\, 
D^{HQ}_{\mu_1 \mu_2...\mu_{\sigma -1};\nu_1\nu_2...\nu_{\sigma-1} } \, .
\label{clstr_prop}
\end{equation}

Let us consider, for concreteness, the case of the spinor-vector $\Psi_\mu $.
Because Proca's equation in (\ref{Dirac_Proca}) eliminates the spin--zero 
(time) component from the $\lbrace 1/2,1/2\rbrace $ representation 
(one Lorentz index)  and ensures that the four--vector
describes a spin-1 field,
the lowest spin-1/2 state will drop out of the
multi--spinor and can be described
independently by the Dirac equation.
With that, the ($S_{2I, 1}, D_{2I, 3}$ ) cluster is now described 
in terms of the Lorentz vector with Dirac spinor components
$\Psi_\mu $ from Eq.~(\ref{class_scheme}) 
and its propagator is given by \cite{Ki97-98a} 
\begin{equation}
S_{\mu\nu} = {{(\gamma\cdot\, p +M)(g_{\mu\nu}
-{1\over M^2} p_\mu p_\nu  )}
\over {2M(p^2-M^2)}}\, .
\label{cluster_prop}
\end{equation}
In noting that, say, the first S$_{11}$ and D$_{13}$ 
states are separated by only 15 MeV, one sees that calculating the relativistic 
contribution of these states to the amplitude of processes like meson 
photoproduction at threshold, is now straightforward. Along the line 
of the  representation theory of the Lorentz group, both the construction of
cluster propagators and interactions with external fields
are also straightforward. For example,
for the case of a ${\cal B}\to N +V$ process, where 
${\cal B}$ stands for a Lorentz covariant spin- and parity cluster, 
while $V$ is a vector meson, a possible effective Lagrangian can be 
written as 
\begin{eqnarray}
{\cal L}_{{\cal B}VN} = 
\bar\Psi ^{\mu_1\mu_2...\mu_{\sigma -1}}
({f_\sigma \over m_\pi^{\sigma -2}}
\partial_{\mu_2}...
\partial_{\mu_{\sigma -1 }}
 A_{\mu_1} &+&  {{f_\sigma '}\over m_\pi^{\sigma -1} }
\partial_{\mu_1} ... \partial_{\mu_{\sigma -1 }} A\!\!\! / )\, 
\psi_N\, , \end{eqnarray}
where $A_\mu $ denotes the vector meson field, while
 $f_\sigma  $ and $f_\sigma ' $ can be fitted to data. 

From the Lorentz--Dirac index notation for the spin- and parity 
clusters in Eq.~(\ref{class_scheme}) one directly reads off that 
the first resonance-cluster  will predominantly couple to 
systems carrying each one Dirac and one Lorentz index
like the pion--nucleon (or $\eta $--nucleon) systems. 
On the contrary, the second and third spin--clusters
will prefer couplings to multipion--nucleon final states
(one Dirac- and several Lorentz indices) in agreement with the empirical 
observations. According to that,
the reason for the observed suppression of the S$_{11}$(1650)$\to $N+$\eta $
decay channel as compared to the S$_{11}$(1535)$\to $N+$\eta $ one,
can be a simple re--distribution of decay strength in favor of the
new opened S$_{11}$(1650)$\to $N+$\pi $+$\pi$ channel.

Now, it is easy to check with the help of Eqs.~(\ref{party}), and
(\ref{coupl_scheme}) that the three different Rarita--Schwinger spinors 
considered by us are distributed over two distinct  Fock spaces,
denoted by ${\cal F}_1$, and ${\cal F}_2$, respectively, which
have opposite vacuum parities and are separated by a well pronounced gap of 
about 300 MeV:
\begin{eqnarray}
{\cal F}_1:\quad 2_{2I ,+} \, :
\quad \Psi_\mu \, &:& P_{2I,1}; S_{2I,1}, 
D_{2I, 3}\, ,
\quad \mbox{for}\quad I= 0,\, {1\over 2},\, {3\over 2}\, ,
\quad \mbox{and}
\nonumber\\
{\cal F}_2:\quad 4_{2I, -} \, :\quad \Psi_{\mu\mu_1\mu_2}\, &:&
S_{2I,1};P_{2I,1}P_{2I,3};D_{2I,3},D_{2I,5}; F_{2I,5}, F_{2I,7}\, ,
\nonumber\\
{\cal F}_2:\quad 6_{2I, -} \, :\quad \Psi_{\mu\mu_1...\mu_4} \, &:&
S_{2I,1};P_{2I,1}P_{2I,3};D_{2I,3},D_{2I,5}; F_{2I,5}, F_{2I,7}\, ;\nonumber\\
&&G_{2I,7}, G_{2I,9};H_{2I,9},H_{2I,11}\, ,\quad
\mbox{for}\quad I= {1\over 2},\,  {3\over 2}\, .
\label{class_scheme}
\end{eqnarray}
The first Fock space, ${\cal F}_1$, is built
upon a $0^+$ vacuum and appears in the spectra of all three baryons
$N$, $\Delta$ and $\Lambda $. It always  contains only the lowest O(4) 
partial wave  with $\sigma =2$ and $\eta =+1$. In contrast to this,
the second Fock space, ${\cal F}_2$, has a $0^-$ vacuum and 
contains the $\sigma$ =4, and 6 partial waves with $\eta$= -1. 
The positive parity of the vacuum 
of ${\cal F}_1$ reflects the realization of chiral symmetry in the hidden 
Nambu--Goldstone mode at low energies. This mode is well known to be 
characterized by a non-vanishing vacuum quark condensate.  
On the contrary, in ${\cal F}_2$ chiral symmetry
must be restored because an isotriplet scalar boson of even 
$G$ parity, as would be required for the Goldstone boson 
of a hidden mode there, is absent from the spectrum. 
The $N\to 4_{1,-}$ and $N\to 6_{1,-}$ excitations
are, therefore, chiral phase transitions. 
Phase transitions of that type have been studied, for example, in 
\cite{Hats,Kim} by means of the change of the quark condensate 
with temperature within the framework of the modified $\sigma $-
model with parity doubling \cite{deTar}.
There, the S$_{11}$(1535) state has been considered as the chiral partner
to the nucleon and the $N\to $S$_{11}$ excitation was shown to reveal the 
hysteresis behavior typical for 1st order phase transitions. 
{}From the considerations given above follows that the S$_{11}$(1535) 
resonance can not be considered as the chiral partner of the nucleon as its
internal orbital angular momentum is $l=1^-$ instead of the required $l=0^-$.
The lightest spin-1/2$^-$ resonance which is built upon a pseudoscalar Fock 
vacuum and satisfies thereby the criteria for a parity partner to the
nucleon, is the second S$_{11}$(1650) state.
It is this resonance that has to enter the calculation of the 
parameters of the chiral phase transitions for baryons.

Finally, special attention has to be paid to the $\Lambda $ hyperon 
excitations, where only the first O(4) partial wave $2_{0,+} $ has been found 
to join the S$_{01}$(1405), D$_{03}$(1520), and the P$_{01}$(1600) states.
Here the mass degeneracy of the resonances is not as well
pronounced as compared to the non--strange sector. In addition,
above 1800 MeV one finds {\bf exact\/} parity degeneracy of states such 
like the S$_{01}$(1800)--P$_{01}$(1810), 
and the P$_{03}$(1890)--D$_{03}$(2000) pairs. A possible interpretation 
of this phenomenon could be the existence of a reflection--asymmetric 
hyperon shape. That such a shape can be the reason for
the occurrence of parity doublets in baryon spectra was considered
repeatedly over the years by different authors \cite{doublets}. 
Such parity pairs can equally well be described relativistically
as they can be mapped onto 
the members of the new type of Lorentz multiplets 
$\lbrace m,0\rbrace \oplus \lbrace 0,m\rbrace
\otimes \lbrack \lbrace {1\over 2},0\rbrace \oplus \lbrace 0, {1\over 2}
\rbrace \rbrack $. 
{}For example, the $\Lambda $ cluster 
$({1\over 2}^+$-${1\over 2}^-$;${3\over 2}^+$-${3\over 2}^-$)
could be identified with the
$\lbrace 1,0\rbrace \oplus \lbrace 0,1\rbrace
\otimes \lbrack \lbrace {1\over 2},0\rbrace \oplus \lbrace 0, {1\over 2}
\rbrace \rbrack $ space which is a totally antisymmetric
2nd rank Lorentz tensor with Dirac spinor components,
$\Psi_{\lbrack \mu ,\nu \rbrack }$ (see \cite{Ki97-98a} for details).

In summary, baryon resonances group to hyperspherical O(4)$_{ls}$
spin--parity clusters rather than to ordinary O(3)$_l$ partial waves,
and the non--strange baryon spectra are completely generated by
the relativistic su(2)$_I\otimes $o(1,3)$_{ls}\otimes $su(3)$_c$ group
algebra rather than by the algebra of the non--relativistic
SU(6)$_{sf}\otimes $O(3)$_l\otimes $SU(3)$_c$ group.
In other words, in performing an O(4) partial wave decomposition of
the $\pi N$ scattering amplitude, one would find the three H\"ohler poles
2$_{2I, +}$, 4$_{2I,-}$, and 6$_{2I, -}$ on the complex energy
plane rather than several dozens independent O(3) partial waves.
This specifics of the $\pi N$ scattering may be related to the role
of the pion as the Goldstone boson of the spontaneously broken 
chiral symmetry SU(2)$_L\otimes $SU(2)$_R$, the space-time version
of which acts as the universal covering of our O(4).
 
The major advantage of our new relativistic spectrum generating 
algebra for baryons is that it reconciles such seemingly 
contrary ideas of the baryon structure like the constituent quark
model, on the one side, and the multi--spinor representations of
the Lorentz group used for structureless particles, on the other side. 
In case, the 'missing' resonances would not be found experimentally,
the canonical belief about the three fermionic degrees of freedom of
baryons has to be revised towards the O(4) symmetric quark-hyperquark 
degrees of freedom, and thereby, towards a new symmetry of strong interaction 
of purely relativistic origin. The O(4) symmetry of the 
internal orbital angular momenta is likely to emerge as the low mass limit
of the approximate conformal SO(2,4) symmetry of the QCD lagrangian 
for the light-flavor quarks.

The scenario of the present work indicates that for the $\pi N$ scattering 
channel, the quark--hyperquark configurations decouple from the remaining 
3q--states. In channels with non--Goldstone mesons, such like, say,
the $\omega N$ one, these couplings may not vanish. In such a case
the baryon excitations would acquire much richer structure and some of the 
traditional `missing' resonances could occur in the spectra 
as supplementary to the ruling O(4) pattern,
an idea originally due to Ref.~\cite{Capstick}.
In the light of our O(4) systematics, 
the success of the traditional U(3) symmetric quark models in describing the 
dynamical properties of the resonances such like their branching ratios, 
form factors etc. is nonetheless understandable in so far that 
the physical observables are, in principle, independent 
on the choice of the Hilbert space basis, provided, the configuration
spaces exploited are large enough.
Our point here is that the O(4) symmetric quark-hyperquark model in
Eq.~(\ref{Hamlit}) is the most economical starting point for resonance 
studies and a serious candidate as a  guiding rule in designing the baryon 
spectra.

Work supported partly by CONACyT Mexico and the Deutsche Forschungsgemeinschaft
(SFB 443).

\begin{figure}[htbp]
\centerline{\psfig{figure=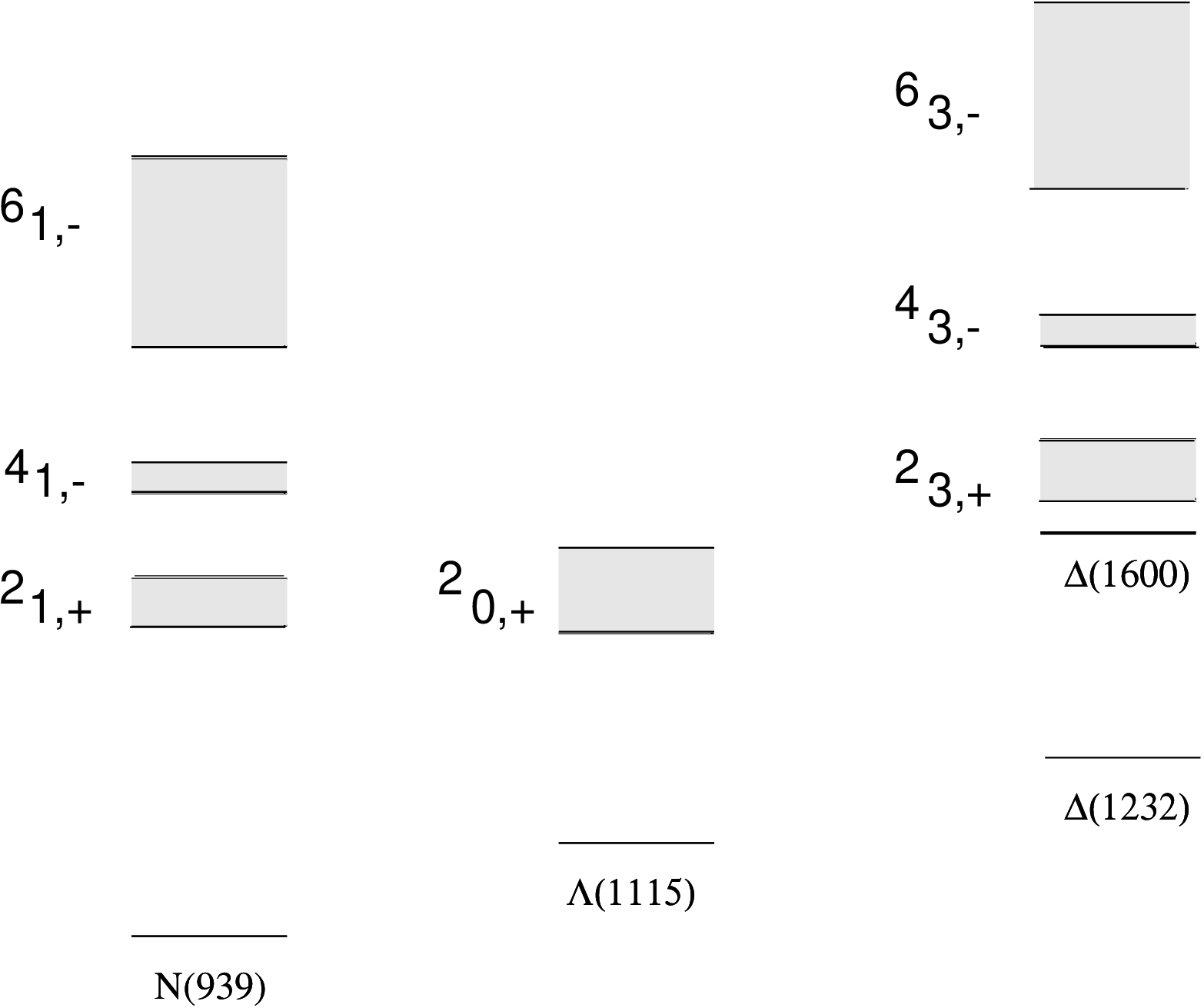,width=10cm}}
\vspace{0.1cm}
{\small Fig. 1\hspace{0.2cm} 
Baryon spectra in terms of the O(4)
partial waves $\sigma_{2I, \eta } $.
Each dashed area contains
$2\sigma -1 $ different spin-states. The appearance of a second isolated
$P_{33}$ state must be of dynamical origin and can contain,
for example, nucleon--gluon components, a possibility considered in
Ref.~ \cite{Barnes}. We expect a second P$_{11}$
resonance paralleling the hybrid $P_{33}$(1600) state in the nucleon sector
to strongly overlap with the Roper resonance from the lowest 
$2_{1, +} $ partial wave and be hidden there.
 }
\end{figure}

\begin{table}[htbp]
\caption{Predicted (M$_\sigma ^{th} $) and reported 
(M$_\sigma ^{ exp} $) positions (in MeV) of the Balmer-like 
baryon lines together with the maximal deviation ($\delta^{ max} $) 
of a resonance mass from the cluster mass-average value. }

\vspace*{0.21truein}

\begin{tabular}{lcccc}
\hline
\\
  $ \sigma_{2I, \eta} $  &  M$_\sigma ^{th}$   & 
  M$_\sigma ^{exp}$  &  $\delta^{max} $& H\"ohler pole \\ \\ 
~\\
\hline
~\\
$2_{1,+}  $ & 1441 & 1498 &58&  \\ \\
$4_{1,-}  $ & 1763&1689& 31& (1665$\pm $25)-i(55$\pm $15) \\ \\
$6_{1, -} $ &2113 &2102& 148&(2110$\pm $50) -i(180$\pm $50)\\
~\\
\hline
~\\
 $2_{3, +} $ &
1734 & 1690&70&  \\ \\
$4_{3, -}$ &
2056& 1922&28 &(1820$\pm $30)-i(120$\pm $30)  \\ \\
$6_{3, -}$ & 
2406 & 2276 &144 &\\ 
~\\
\hline
~\\
 $2_{0, +}  $  & 
 1618&1508& 103\\
~\\
\end{tabular}
\end{table}

\end{document}